УДК 539.194

# О НАХОЖДЕНИИ ЭЛЕКТРОННО-КОЛЕБАТЕЛЬНО-ВРАЩАТЕЛЬНЫХ ТЕРМОВ ДВУХАТОМНЫХ МОЛЕКУЛ ПО ИЗМЕРЕННЫМ ВОЛНОВЫМ ЧИСЛАМ

Б. П. Лавров и М. С. Рязанов

*Физический факультет, Санкт-Петербургский государственный университет,*
*198904 Петергоф, Санкт-Петербург, Россия*
*e-mail: lavrov@pobox.spbu.ru*

## АННОТАЦИЯ

Предлагается методика нахождения электронно-колебательно-вращательных (ЭКВ) термов двухатомных молекул из экспериментальных данных о волновых числах ЭКВ спектральных линий. В отличие от существующих, она основана только на комбинационном принципе Ридберга–Ритца. Показано, что связь между совокупностями абсолютных значений ЭКВ термов и волновых чисел наблюдаемых спектральных линий появляется при наличии трёх и более ЭК состояний, попарно связанных радиационными переходами. Новый метод отличается от существующих тем, что он: 1) не требует никаких дополнительных предположений о строении молекулы; 2) не нуждается в применении промежуточных параметров, таких как молекулярные константы в традиционных методах; 3) позволяет в рамках одноступенчатой оптимизационной процедуры использовать все имеющиеся в наличии экспериментальные данные, полученные для разных систем полос разными авторами и в разных работах; 4) в интерактивном режиме даёт возможность проводить рациональную селекцию экспериментальных данных и исключать промахи, исправлять ошибки идентификации ЭКВ спектральных линий и проводить анализ согласованности различных наборов данных; 5) позволяет получить не только оптимальные значения ЭКВ термов, но и погрешности их определения, обусловленные объёмом и качеством имеющихся экспериментальных данных. Предлагаемая методика основана на минимизации среднеквадратичного отклонения измеренных и вычисленных значений волновых чисел ЭКВ спектральных линий. В силу линейности используемых уравнений эта оптимизационная задача сводится к решению системы линейных алгебраических уравнений. Необходимым условием для применения данного метода является предварительная идентификация линий ЭКВ спектра, то есть нахождения соответствия наблюдаемых ЭКВ линий и конкретных ЭКВ переходов. Для этой цели сохраняется необходимость использования традиционных методов анализа спектров.

## 1. ВВЕДЕНИЕ

Одной из важнейших характеристик молекулы, как и любой квантовой системы, является спектр собственных значений энергии, то есть совокупность электронно-колебательно-вращательных (ЭКВ) термов. Термы связаны с наблюдаемыми на опыте волновыми числами ЭКВ спектральных линий испускания и поглощения простым соотношением — комбинационным принципом Ридберга–Ритца (соответствующем правилу частот Бора):

$$\nu_{nvJ}^{n'v'J'} = T_{n'v'J'} - T_{nvJ}, \tag{1}$$

где $\nu_{nvJ}^{n'v'J'}$ — волновое число[1] линии излучения или поглощения при переходе $n'v'J' \to nvJ$, через $n$ обозначены наборы квантовых чисел, определяющие электронные состояния, $v$ и $J$ — колебательные и вращательные квантовые числа[2]. Видно, что измеримые на опыте волновые числа спектральных линий непосредственно связаны только с разностями ЭКВ термов, поэтому абсолютные значения термов не могут быть найдены таким образом напрямую. Существующие в настоящее время способы экспериментального нахождения термов основаны на принципе Ридберга—Ритца, но требуют внесения некоторых дополнительных предположений о свойствах молекулы.

---

[1]Здесь и далее используются волновые числа, отнесённые к вакууму.

[2]Для простоты ограничимся рассмотрением синглетных термов.

В классическом методе [1] из измеряемых волновых чисел находят не сами термы, а так называемые молекулярные константы, которые являются коэффициентами разложения зависимости энергии от квантовых чисел в некоторые ряды, в частности, широко применяемого интерполяционного многочлена (разложения Данхэма [2]):

$$T_{nvJ} = \sum_{kl} Y_{kl}(v+\tfrac{1}{2})^k [J(J+1)]^l. \qquad (2)$$

Обычно коэффициенты разложения быстро убывают с ростом $k$ и $l$, и можно ограничиться лишь несколькими первыми членами ряда. В спектроскопии его часто записываю в виде

$$\begin{aligned}
T_{nvJ} &= T_v + B_v J(J+1) - D_v [J(J+1)]^2 + \ldots, \\
T_v &= T_e + \omega_e(v+\tfrac{1}{2}) - \omega_e x_e (v+\tfrac{1}{2})^2 + \ldots, \\
B_v &= B_e - \alpha_e(v+\tfrac{1}{2}) + \ldots, \\
D_v &= D_e - \beta_e(v+\tfrac{1}{2}) + \ldots.
\end{aligned} \qquad (3)$$

При использовании этого метода идентификация спектра производится одновременно с нахождением данных коэффициентов для анализируемого электронного или электронно-колебательного перехода (системы полос или отдельной полосы соответственно). Например, вращательные константы находятся из так называемых комбинационных разностей (см. рис. 1), дающих разности ЭКВ термов только одного ЭК состояния — нижнего:

$$\nu^{n'v'J}_{nvJ-1} - \nu^{n'v'J}_{nvJ+1} = T_{nvJ+1} - T_{nvJ-1} = (4B_v - 6D_v)(J+\tfrac{1}{2}) - 8D_v(J+\tfrac{1}{2})^3 + \ldots \qquad (4)$$

или верхнего:

$$\nu^{n'v'J+1}_{nvJ} - \nu^{n'v'J-1}_{nvJ} = T_{n'v'J+1} - T_{n'v'J-1} = (4B'_v - 6D'_v)(J+\tfrac{1}{2}) - 8D'_v(J+\tfrac{1}{2})^3 + \ldots \qquad (5)$$

Закономерное поведение зависимостей комбинационных разностей от вращательных и колебательных квантовых чисел используется в качестве критерия правильности идентификации спектральных линий. После нахождения вращательных констант различных ЭК состояний последовательно находятся также колебательные, и затем электронные константы.

Достоинства метода заключаются в том, что он позволяет производить анализ ещё не идентифицированных спектров, прост в реализации и даёт компактную форму представления результатов (сотни ЭКВ термов могут быть представлены десятком молекулярных констант).

Существенными недостатками являются:

1. Полиномиальные разложения при разумном количестве коэффициентов не способны описать нерегулярные возмущения. Более сложные формулы теории возмущений позволяют улучшить описание, но приводят к потере компактности и простоты использования.

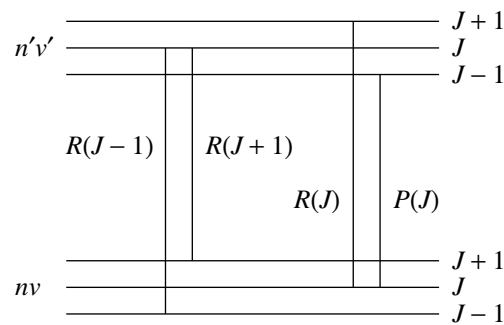

Рис. 1. Диаграмма Гротриана, поясняющая возникновение комбинационных разностей (5) и (4).

2. Отдельные полосы и системы полос анализируются раздельно. Это не позволяет получить оптимального набора констант. Часто константы для одного и того же электронного состояния, полученные по разным системам полос, не совпадают в пределах погрешностей.

3. Зависимость термов от квантовых чисел ищется в виде разложения в определённый ряд, то есть получаемые значения молекулярных констант зависят от выбора самого вида ряда (выбора модели).

4. Разложение ведётся по неортогональным функциям, поэтому получаемые значения констант зависят от выбора длины ряда.

5. Набор молекулярных констант зависит от количества описываемых значений термов и от весов, с которыми брались волновые числа. Ни то, ни другое обычно не указывается. Указываются также лишь погрешности нахождения самих констант (и то не всегда!), но не точность, с которой они описывают значения термов.

В работе [3] было предложено одновременного находить константы для нескольких электронных состояний, а в [4] предложена и реализована одноступенчатая процедура нахождения полного оптимального (для данного набора экспериментальных данных и выбранной модели) набора констант. Процедура позволяет получить и более грамотные оценки погрешностей. Метод заключается в минимизации функционала невязки[3]:

$$\Phi = \sum \left[ (T_{n'v'J'} - T_{nvJ}) - \nu_{nvJ}^{n'v'J'} \right]^2,$$

где $T_{nvJ}$ представлены в виде рядов (2). Это позволяет найти оптимальный набор параметров $Y_{kl}$ для всех электронных состояний и матрицу ковариации, диагональные элементы которой предлагалось использовать в качестве оценок погрешностей получаемых констант.

Метод полностью устраняет 2-й недостаток классического метода и серьёзно уменьшает 5-й, наследуя при этом остальные недостатки.

Следует сделать несколько замечаний об особенностях любых методов, основанных на получении информации о термах через молекулярные константы.

1. Проблема выбора длины ряда. Очевидно, что чем большее число членов используется, тем точнее получается описание экспериментальных данных. Но известно также, что подобные многочлены имеют свойство «раскачиваться» на краях. Получается некое противоречие: чем лучше описание, тем хуже предсказание. В некоторых случаях для приемлемого описания может понадобиться количество констант, почти равное количеству описываемых термов (например, в [22] потребовалось 6 и 5 колебательных констант для описания 7 и 8 колебательных уровней). При этом исчезает преимущество компактного описания и совершенно теряется физический смысл получаемых величин.

2. Связанная с предыдущей и количеством данных. Как уже обсуждалось в [4], в приложениях следует использовать константы, найденные для того же количества уровней и с использованием той же модели (вида и длины ряда). Отбрасывать коэффициенты при высоких степенях нельзя из-за неортогональности. Использование констант, описывающих большее число уровней, приводит к неоптимальности (систематическим ошибкам), описывающих меньшее — к большим непредсказуемым ошибкам для верхних уровней.

3. Проблема погрешностей. Молекулярные константы ищутся для оптимального описания экспериментальных значений волновых чисел. Из-за неполной адекватности модели при этом в описании термов могут появляться более или менее существенные систематические отклонения, величину которых определить невозможно. Для оценки погрешностей производных от констант величин (в т.ч. термов) можно использовать матрицу ковариации, но, как отмечается в [4], получаемые величины описывают скорее точность аппроксимации, т.е. систематические отклонения. В работах также часто приводятся разницы «O-C» измеренных и рассчитанных по константам волновых чисел, позволяющие оценить качество их описания. При этом авторы иногда (например, в [21]) совершенно не обращают внимания на то, что практически все разности получились одного знака!

---

[3]В [4] вместо волновых чисел использовались значения термов, полученные в [5], но говорилось о возможности непосредственного использования волновых чисел.

Недостатки предыдущих методов заставили авторов работ [5] и [6] находить термы, непосредственно используя комбинационные разности. Если положить энергию основного ЭКВ состояния равной нулю, то из (4) можно найти энергию второго вращательного уровня. Далее можно найти энергию уровня с $J = 4$ и т.д. для всех чётных $J$. Аналогично можно выстроить уровни с нечётным $J$ относительно первого вращательного уровня. Связать уровни с чётным и нечётным $J$ между собой таким образом невозможно, поскольку, согласно правилу Лапорта, переходы возможны только между ЭКВ уровнями разной чётности ($+$ и $-$), и после двух переходов чётность не меняется, а соседние вращательные уровни имеют разные чётности. Для решения данной проблемы использовались дополнительные предположения:

1. Низкие вращательные термы обычно хорошо описываются рядом вида (3), поэтому, найдя вращательные константы только по нескольким нижним чётным термам, можно рассчитать по ним энергию первого уровня и связать таким образом нечётные с чётными.

2. В некоторых электронных состояниях с $\Lambda \neq 0$ на нижних вращательных уровнях $\Lambda$-удвоение достаточно мало. Если в таких случаях пренебречь этой разницей, то предположение, что энергии нижних уровней $\Lambda$-дублета совпадают, также позволяет решить проблему. Далее последовательно выстраивались термы других электронно-колебательных состояний.

Достоинствами метода являются:

1. Резкое уменьшение зависимости конечного результата от используемого вида представления, поскольку ряд (2) на низких уровнях работает достаточно хорошо, и ошибку в определении энергии первого уровня можно сделать малой.

2. Оценка погрешностей существенно облегчается. Фактически, остаётся только экспериментальная погрешность. В [6] для контроля использовалось построение гистограмм невязок (разностей измеренных и вычисленных из полученных термов волновых чисел), хотя сами погрешности термов не приведены.

3. Последующее расширение экспериментальных данных о волновых числах линий, связанных с более высокими вращательными уровнями не влияет на полученные значения термов более низких вращательных уровней.

Среди недостатков надо отметить необходимость дополнительных (кроме принципа Ридберга—Ритца) предположений, а также многоступенчатость процедуры, из которой следует неоптимальность: в частности, ошибка (промах) в определении какого-либо ЭКВ терма приводит к систематическому сдвигу всех вышележащих, вычисленных с его помощью.

В работе [7] (см. также [8]) для определения термов на основе комбинационного принципа Ридберга—Ритца было предложено использовать решение переопределённой системы уравнений (1) с помощью ЭВМ. Метод является принципиально одноступенчатым, но в силу ограниченных возможностей ЭВМ того времени приходилось ограничиваться обработкой только одного электронного состояния. Для нахождения относительного положения некомбинирующих уровней по-прежнему использовалась полиномиальная аппроксимация.

Цель настоящей работы состояла в разработке методики определения набора ЭКВ термов из всей имеющейся совокупности экспериментально измеренных волновых чисел, не требующей дополнительных предположений о свойствах молекулы, но основанной только на принципе Ридберга—Ритца, справедливость и общность которого не вызывает сомнений. Создание такой методики позволило бы не только получать наилучший набор термов, но и проводить анализ имеющегося материала о волновых числах ЭКВ спектральных линий, полученного на разных экспериментальных установках, разными авторами и в разное время, а следовательно, производить рациональный (объективный) отбор наиболее достоверных экспериментальных данных.

## 2. ОПИСАНИЕ МЕТОДА

Предлагаемый метод предполагает, что ЭКВ спектр молекулы уже идентифицирован (т.е. наблюдаемые спектральные линий отождествлены конкретным ЭКВ переходам), поэтому его можно считать дополнительным по отношению к классическому процессу использования комбинационных разностей. В дальнейшем будем считать, что имеющаяся идентификация является однозначной и достоверной (для этого необходима информация не только о волновых числах, но и об интенсивностях — см. [9]).

Основная идея метода заключается в том, что при одновременном использовании экспериментальных данных для всех систем полос в системе уравнений (1), в принципе, содержится информация об абсолютных значениях всех ЭКВ термов. В частности, для разности энергий соседних вращательных уровней можно использовать соотношения волновых чисел не двух, как в традиционном методе, а трёх спектральных линий (см. рис. 2):

$$T_{nvJ+1} - T_{nvJ} = \nu_{nvJ}^{n'v'J+1} + \nu_{n'v'J+1}^{n''v''J+2} - \nu_{nvJ+1}^{n''v''J+2}, \quad (6)$$

$$T_{nvJ+1} - T_{nvJ} = \nu_{nvJ}^{n''v''J+1} - \nu_{n'v'J}^{n''v''J+1} - \nu_{nvJ+1}^{n'v'J} \quad (7)$$

и т.п. Так, соотношению (6) в системе уравнений (1) соответствуют следующие уравнения:

$$\begin{cases} \nu_{nvJ}^{n'v'J+1} = T_{n'v'J+1} - T_{nvJ} \\ \nu_{n'v'J+1}^{n''v''J+2} = T_{n''v''J+2} - T_{n'v'J+1} \\ \nu_{nvJ+1}^{n''v''J+2} = T_{n''v''J+2} - T_{nvJ+1} \end{cases},$$

или, в матричной форме,

$$\begin{pmatrix} -1 & 0 & 1 & 0 \\ 0 & 0 & -1 & 1 \\ 0 & -1 & 0 & 1 \end{pmatrix} \begin{pmatrix} T_{nvJ} \\ T_{nvJ+1} \\ T_{n'v'J+1} \\ T_{n''v''J+2} \end{pmatrix} = \begin{pmatrix} \nu_{nvJ}^{n'v'J+1} \\ \nu_{n'v'J+1}^{n''v''J+2} \\ \nu_{nvJ+1}^{n''v''J+2} \end{pmatrix}. \quad (8)$$

После приведения к диагональному виду получаем:

$$\begin{pmatrix} -1 & 1 & 0 & 0 \\ -1 & 0 & 1 & 0 \\ -1 & 0 & 0 & 1 \end{pmatrix} \begin{pmatrix} T_{nvJ} \\ T_{nvJ+1} \\ T_{n'v'J+1} \\ T_{n''v''J+2} \end{pmatrix} = \begin{pmatrix} \nu_{nvJ}^{n'v'J+1} + \nu_{n'v'J+1}^{n''v''J+2} - \nu_{nvJ+1}^{n''v''J+2} \\ \nu_{nvJ}^{n'v'J+1} \\ \nu_{nvJ}^{n'v'J+1} + \nu_{n'v'J+1}^{n''v''J+2} \end{pmatrix},$$

первая строка чего и даёт искомое соотношение (6).

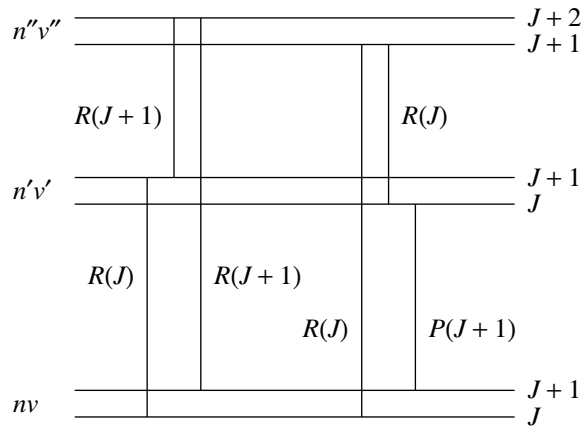

Рис. 2. Диаграмма Гротриана, поясняющая возникновение соотношений (6) и (7).

Таким образом, решая систему уравнений (1), составленную для всех имеющихся экспериментальных данных по всем системам полос, можно найти все участвующие в ней термы без каких-либо дополнительных предположений. Система является переопределённой (это видно уже из пары уравнений (6) и (7)) и, следовательно, несовместной, поскольку экспериментальные данные всегда содержат ошибки измерений. Для нахождения оптимального набора ЭКВ термов можно воспользоваться стандартным методом (описанным, например, в [10]) минимизации квадрата невязки

$$r^2 = \sum_{\nu_{nvJ}^{n'v'J'}} \left[ \frac{(T_{n'v'J'} - T_{nvJ}) - \nu_{nvJ}^{n'v'J'}}{\sigma_{\nu_{nvJ}^{n'v'J'}}} \right]^2, \qquad (9)$$

где $\sigma_{\nu_{nvJ}^{n'v'J'}}$ — оценки погрешностей измеренных волновых чисел, а сумма берётся по всем имеющимся экспериментальным данным. Если экспериментальные ошибки распределены по нормальному закону, метод наименьших квадратов приводит к решению, отвечающему принципу максимального правдоподобия.

Классификация термов по индексам $n$, $v$ и $J$ важна при их использовании и сопоставлении наблюдаемых волновых чисел ЭКВ уровням, между которыми происходят переходы, в частности, для применения правил отбора. Когда такое сопоставление проведено, конкретные обозначения термов и линий уже не важны. Фактически, можно перенумеровать интересующие термы натуральными числами в произвольном порядке, перейдя от обозначений $T_{nvJ}$ к более компактным $T_i$, а волновые числа обозначать парой индексов, соответствующих начальному и конечному уровням:

$$\nu_{ij} = T_i - T_j. \qquad (10)$$

Если объём экспериментальных данных составляет $n_\nu$ волновых чисел, а количество искомых термов равно $n_T$, то систему уравнений (10) можно записать в матричной форме в виде

$$\mathbf{N} = \mathbf{AT} + \mathbf{R},$$

где $\mathbf{N}$ — вектор экспериментальных волновых чисел, $\mathbf{T}$ — вектор термов, $\mathbf{R}$ — вектор остатков, а $\mathbf{A}$ — структурная матрица размерности $n_\nu \times n_T$, подобная матрице в выражении (8). Погрешности волновых чисел описываются их матрицей ковариации $\mathbf{D}(\nu)$ размерности $n_\nu \times n_\nu$. Тогда квадрат невязки можно выразить как $r^2 = \mathbf{R}'\mathbf{D}(\nu)\mathbf{R}$, где штрих означает транспонирование. Поскольку корреляции волновых чисел неизвестны, будем считать матрицу $\mathbf{D}(\nu)$ диагональной с элементами, равными квадратам экспериментальных погрешностей соответствующих волновых чисел, что соответствует выражению (9). Условие минимума невязки приводит к системе нормальных уравнений

$$\mathbf{A}'\mathbf{D}^{-1}(\nu)\mathbf{AT} = \mathbf{A}'\mathbf{D}^{-1}(\nu)\mathbf{N}. \qquad (11)$$

Обозначим $\mathbf{C} \equiv \mathbf{A}'\mathbf{D}^{-1}(\nu)$, $\mathbf{B} \equiv \mathbf{CA}$. Тогда (11) перепишется как

$$\mathbf{BT} = \mathbf{CN}. \qquad (12)$$

Диагональность $\mathbf{D}(\nu)$ позволяет написать матрицу $\mathbf{B}$ и вектор $\mathbf{CN}$ в явном виде:

$$\mathbf{B} = \begin{pmatrix} \sum_k w_{1k} & -w_{12} & -w_{13} & \ldots & -w_{1n_T} \\ -w_{12} & \sum_k w_{2k} & -w_{23} & \ldots & -w_{2n_T} \\ \multicolumn{5}{c}{\dotfill} \\ -w_{1n_T} & -w_{2n_T} & -w_{3n_T} & \ldots & \sum_k w_{n_T k} \end{pmatrix}, \quad \mathbf{CN} = \begin{pmatrix} \sum_k w_{1k}\nu_{1k} \\ \sum_k w_{2k}\nu_{2k} \\ \ldots \\ \sum_k w_{n_T k}\nu_{n_T k} \end{pmatrix}.$$

Здесь введено обозначение $w_{ij} = 1/\sigma^2_{\nu_{ij}}$ для весов (для отсутствующих переходов $w_{ij} = 0$), и считаем $w_{ji} = w_{ij}$, $\nu_{ji} = -\nu_{ij}$. Если какое-либо волновое число было измерено в нескольких экспериментах, оно учитывается соответствующее число раз, для этого следует заменить $w_{ik}$ на $w_{ik}^{(1)} + w_{ik}^{(2)} + \ldots$ в левой части и $w_{ik}\nu_{ik}$ на $w_{ik}^{(1)}\nu_{ik}^{(1)} + w_{ik}^{(2)}\nu_{ik}^{(2)} + \ldots$ в правой. Как видно, $\mathbf{B}$ получается вырожденной (сумма всех строк равна нулю), что соответствует произвольному выбору начала отсчёта энергии. Для получения однозначного решения нужно фиксировать значение какого-либо терма $T_i = T_i^{fix}$. Это осуществляется заменой $i$-й строки матрицы строкой $(0, \ldots, 0, 1, 0, \ldots, 0)$ с единицей в $i$-м столбце и заменой $i$-й компоненты вектора $\mathbf{CN}$ на $T_i^{fix}$. Если имеется несколько не связанных между собой наборов термов, например, разной мультиплетности, то таким образом фиксируются начала отсчёта в каждой группе.

Для оценки матрицы ковариации найденных из решения системы (12) ЭКВ термов имеем:

$$\hat{\mathbf{D}}(T) = \frac{r^2}{n_\nu - n_T + n_T^{fix}} \mathbf{B}^{-1},$$

где $n_T^{fix}$ — количество фиксированных термов. Если оценки погрешностей измерений волновых чисел соответствуют действительной точности эксперимента, дробь перед $\mathbf{B}^{-1}$ должна быть близка к единице.

Таким образом, для полного решения поставленной задачи достаточно иметь возможность обращать достаточно разреженные матрицы размерности $n_T \times n_T$. Для $n_T$ порядка тысячи, как обычно и бывает, это можно делать даже на современных персональных компьютерах.

После нахождения значений термов следует проанализировать несмещённые оценки отклонений
$$R'_{ij} = \nu_{ij} - (T'_i - T'_j),$$
где $T'_i$ — значения термов, полученные без учёта данного волнового числа $\nu_{ij}$, которые могут быть выражены через значения термов, полученные с использованием всех волновых чисел, и матрицу ковариации этих термов. Для анализа отклонений удобно построить график функции распределения $F(\xi)$ величины

$$\xi_{ij} = \frac{R'_{ij}}{\sigma_{ij}} = \frac{\nu_{ij} - (T_i - T_j)}{\sqrt{\sigma^2_{ij} - (D_{ii}(T) + D_{jj}(T) - 2D_{ij}(T))}},$$

которое в благоприятном случае должно быть близко к нормальному с нулевым средним и единичной дисперсией. Использование функции распределения вместо гистограмм невязок более информативно, поскольку позволяет работать с выборками любого размера и лучше оценивать как сам вид распределения, так и его параметры, особенно при применении спрямляющего преобразования (как на вероятностной бумаге). Для анализа могут быть полезны выборки, содержащие волновые числа линий:
- всех;
- полученных в определённом эксперименте;
- относящихся к определённой системе полос или полосе;
- относящихся к определённому электронному или ЭК состоянию или ЭКВ терму.

Они позволяют проверять качество соответствующих частей экспериментальных данных и их согласованность между собой, что позволяет производить рациональный отбор материала, отсеивая как отдельные промахи в волновых числах, так и более крупные совокупности — вплоть до данных отдельных экспериментов.

Полученные в результате этой процедуры значения термов с погрешностями и их матрица ковариации могут быть использованы для получения любых других, зависимых от термов, параметров молекулы. Причём, помимо матрицы ковариации, для нахождения этих параметров следует использовать некоторые дополнительные веса для термов, определяемые конкретной задачей.

## 3. ЗАКЛЮЧЕНИЕ

Предлагаемый в настоящей работе метод позволяет получить набор экспериментальных значений ЭКВ термов, не связанный с какими-либо представлениями о строении молекулы[4] и являющийся оптимальным (а значит и наилучшим из возможных) как с точки зрения принципа максимального правдоподобия, так и в отношении имеющегося экспериментального материала. Получаемые значения термов и их погрешности целиком определяются количеством имеющихся экспериментальных данных[5] и точностью измерений. Поэтому найденные значения ЭКВ термов можно считать вторичными экспериментальными данными.

По возможности наиболее точное экспериментальное определение спектра собственных значений двухатомных молекул (набора ЭКВ термов) представляет интерес как с точки зрения квантовой механики молекул (построение модельных гамильтонианов, возможность прямого сравнения с экспериментом результатов «ab initio» расчётов и пр.[6]), так и в прикладной спектроскопии для сравнения наблюдаемых спектров с синтезируемыми на основе тех или иных моделей физических процессов в горячих газах и плазме.

Следует отметить интересную особенность данного метода, существенно отличающую его от методов, основанных на использовании молекулярных констант. Она касается ситуации, когда уже после проведения анализа спектра и получения окончательных результатов в виде набора ЭКВ термов или молекулярных констант некоторого электронного состояния, появится новая работа, в которой будут измерены волновые числа спектральных линий идущих с более высоких колебательных или более высоких вращательных уровней.[7] Такое расширение экспериментального материала не изменит значения ЭКВ термов, полученных ранее с помощью данного метода, а лишь добавит новые, ранее не известные термы. При возникновении той же ситуации в традиционных методах требуется заново пересчитывать весь набор молекулярных констант.

Выше для простоты изложения рассматривался случай синглетных термов гетероядерной молекулы. Понятно, что в случае мультиплетных термов всё изложенное остаётся справедливым, за исключением того, что в классификацию самих термов добавится ещё одно квантовое число и соответствующие правила отбора, а при переходе от формы записи (1) к (10) нужно будет использовать другую перенумерацию. Сложность может возникнуть только при отсутствии исследованных переходов между термами разной мультиплетности — в таком случае соответствующие наборы термов останутся несвязанными, т.е. будут иметь различные начала отсчёта.

В случае гомоядерных молекул возникает принципиальная проблема, связанная с тем, что вращательные уровни с разной чётностью имеют разную симметрию относительно перестановки ядер. Переходы между состояниями различной ядерной симметрии запрещены (запрет не является строгим, но очень сильным, т.к. связан с изменением ориентации ядерного спина), поэтому совокупность ЭКВ термов распадается на две не связанных между собой части (например, уровни орто- и параводорода). Если найти разность между хотя бы двумя термами различной симметрии не удаётся экспериментально, то в этом случае остаётся необходимость

---

[4] Как уже говорилось, использование принципа Ридберга–Ритца не снижает общности в силу его фундаментальности.

[5] На конечный результат влияет также и распределение экспериментальных данных по исследованным колебательным и вращательным уровням в различных электронных состояниях.

[6] В настоящее время результаты «ab initio» расчетов обычно сравниваются с найденными полуэмпирически потенциальными кривыми (см. [11]) либо с отдельными молекулярными константами, в основном, с $T_e$, $\omega_e$ и $B_e$. Последняя однозначно связана с так называемым равновесным межъядерным расстоянием $r_e$. Принимая во внимание условность молекулярных констант (см. выше и [4]), такие сравнения нельзя признать ни прямыми ни полными.

[7] Здесь предполагается, что линии, идущие с более низких уровней ранее были измерены с достаточно высокой точностью. Понятно, что если в новой работе будет доказано, что в предыдущих были допущены ошибки, пересчету подлежат как молекулярные константы, так и ЭКВ термы.

использования каких-либо дополнительных предположений, подобных применённым в работах [5, 6].

На первый взгляд, недостатком предлагаемого метода является относительно большой объём выходных данных (сотни–тысячи значений ЭКВ термов вместо десятков–сотен молекулярных констант при традиционном подходе к проблеме). Однако, вряд ли это можно считать существенным по двум причинам. Во-первых, очевидно, что количество находимых значений ЭКВ термов в любом случае не превосходит (а обычно в несколько раз меньше) количества использованных волновых чисел спектральных линий, таблицы которых, как правило, публикуются в оригинальных работах. Во-вторых, при современном уровне вычислительной техники, баз данных и сетей ни хранение, ни передача, ни обработка подобных объёмов информации не вызывает каких-либо сложностей. В то же время, выходные данные метода, предлагаемого в настоящей работе, несут всю доступную информацию об ЭКВ термах, и любое сокращение их объёма приведёт к потере той или иной её части.

Заметим, что математический формализм, на котором основан данный метод, не содержит никаких особенностей ни двухатомной молекулы, ни молекулы вообще. Специфика двухатомной молекулы здесь заключается лишь в сопоставлении волновых чисел спектральных линий и термов (в частности, в правилах отбора). Это означает, что предлагаемый подход к проблеме, в принципе, может быть применён также и для определения термов как многоатомных молекул, так и атомов.

Следует ещё раз подчеркнуть, что необходимым условием для применения данного метода является предварительная идентификация линий ЭКВ спектра. Для этой цели сохраняется необходимость использования стандартных методов анализа молекулярных спектров, поэтому предлагаемый метод является дополнительным по отношению к традиционным.

Настоящий метод был успешно применён для нахождения ЭКВ термов всех экспериментально исследованных к настоящему времени синглетных электронных состояний изотопомера молекулы гидрида бора $^{11}$BH, а именно: $X^1\Sigma^+$, $A^1\Pi$, $C'^1\Delta$, $B^1\Sigma^+$, $C^1\Sigma^+$, $D^1\Pi$, $E^1\Sigma^+$, $F^1\Sigma^+$, $G^1\Pi$, $H^1\Delta$, $I^1\Sigma^+$, $J^1\Sigma^+$. Были использованы все известные авторам экспериментальные данные о волновых числах ЭКВ спектральных линий, принадлежащих 15 системам полос, полученные в 10 различных работах [12–21]. Анализ данных, проведённый описанным выше образом, показал, что результаты различных работ находятся в достаточно хорошем согласии друг с другом, за исключением ранних работ [12] и [14], в которых обнаружились систематические ошибки порядка 0.05...0.10 см$^{-1}$. Все данные этих двух работ были исключены из дальнейшего рассмотрения. Из оставшихся 1410 волновых чисел 32 числа были исключены как ошибочные по критерию Шовене: $|\xi_{ij}| > 3$. В результате были получены 529 значений ЭКВ термов 12 электронных состояний молекулы BH. Значения ЭКВ термов с погрешностями представлены в [23]. Подробное описание этого опыта применения метода выходит за рамки настоящей статьи и является предметом для отдельной публикации.